# The small-world phenomenon: a model, explanations, characterizations and examples

Leo Egghe[1] and Ronald Rousseau[2,3]

[1]Hasselt University, Hasselt, Belgium

E-mail: leo.egghe@uhasselt.be

ORCID: 0000-0001-8419-2932

[2]KU Leuven, MSI, Facultair Onderzoekscentrum ECOOM,

Naamsestraat 61, 3000 Leuven, Belgium

E-mail: ronald.rousseau@kuleuven.be

&

[3]University of Antwerp, Faculty of Social Sciences,

Middelheimlaan 1, 2020 Antwerp, Belgium

E-mail: ronald.rousseau@uantwerpen.be

ORCID: 0000-0002-3252-2538

Abstract

We introduce and define three types of small worlds: small worlds based on the diameter of the network (SWD), those based on the average geodesic distance between nodes (SWA), and those based on the median geodesic distance (SWMd). These types of networks are defined as limiting properties of sequences of sets. We show the exact relation between these three types, namely that each SWD network is also an SWA network and that each SWA network is also an SWMd network. Yet, having the small-world property is rather evident, in the sense that most networks are small-world networks in one of the three ways. We introduce sequences of distance frequencies, so-called alpha-sequences, and prove a relation between the majorization property between alpha-sequences and small-world properties.

## 1. Introduction

In this article, we will show that the occurrence of the small-world property (precisely defined further on) is rather evident. In this, we fully agree with Newman (2003). According to the original formulation of the small world phenomenon by the Hungarian writer Frigyes Karinthy (1929), people are linked through an acquaintances chain of at most five links. Hence the original case is actually "five degrees of separation" and not "six degrees". We owe this observation to Barabási (2002), who got it from Tibor Braun (see also (Braun, 2004)). When trying to confirm this statement in the real world, as Milgram (1967) and Albert et al. (1999) did, one can only observe that the obtained distance is a given number, but it is very difficult, maybe almost impossible, to find the actual shortest distance between two persons, or in general between two nodes in a network (Killworth et al., 2006). Of course, in theoretical work, this restriction does not apply. Several scientists, see e.g., Barabási (2002), Jackson & Rogers (2007), have pointed out that the number 6 is just a symbol referring to the idea of a small world. It can be replaced by any fixed positive constant K. Yet, Samoylenko et al. (2023) showed that in a specific game-theoretic model the diameter is exactly equal to 6.

The best-known and most-studied small worlds have an additional property, namely they have a high clustering coefficient. In our study, we do not consider this extra property.

## 2. Basic definitions

### 2.1 A distance function

Given a non-empty set $\Omega$, then a distance function (or metric) on $\Omega$ is a mapping d to the non-negative real numbers: $d\colon \Omega \times \Omega \to \mathbb{R}^+$ satisfying the following three conditions:

1) $d(s,t) = 0 \iff s = t$

2) $\forall s, t \in \Omega\colon d(s,t) = d(t,s)$

3) The triangle inequality: $\forall s, t, u \in \Omega\colon d(s,t) \le d(s,u) + d(u,t)$

A set $\Omega$ given together with a distance function d, is called a metric space. A connected graph (or network) with a finite number of nodes is an example of a metric space in which the distance between two nodes is the length of a shortest path, also called a geodesic, between these two nodes.

### 2.2 Diameter of a subset of a metric space

The diameter of a non-empty subset S in a metric space M, denoted as diam(S) is the supremum of the distances between its points. A set is called bounded if it has a finite diameter.

## 2.3 Diameter of a graph (Wasserman & Faust, 1994, p.111)

The diameter of a finite, connected graph is the length of the largest geodesic between any pair of nodes. We note that some scientists such as Dorogovtsev & Mendes (2002) define the diameter of a graph as the average distance between any two connected nodes. We do not follow them and stick to the standard Wasserman-Faust definition.

As real-world networks are often dynamic (the world population changes, for instance) we will work within the context of a sequence of finite sets $(\Omega_N)_{N \in \mathbb{N}}$ , i.e., $\forall\, N \in \mathbb{N}$: $\#\Omega_N < +\infty$ but $\lim_{N \to \infty}\big(\#(\Omega_N)\big) = +\infty$ is allowed. Further on we will need an infinite index set which is not equal to the set of all natural numbers, but consists of an infinite subset of the natural numbers. From a number-theoretic point of view, this amounts to the same as both index sets have the same cardinality (there exists a bijection between the two sets).

Before formulating the notion of a small-world we recall the definition of a multiset (sometimes also called "a family").

## 2.3 Multiset

The notion of a multiset generalizes the notion of a set (Jouannaud & Lescanne, 1982; Rousseau et al., 2018, p. 141). In a multiset, an element can occur several times. A multiset containing the elements x,x,y,z,z,z will be denoted as {{x,x,y,z,z,z}}. As for sets, the order in which the elements of a multiset are shown plays no role.

Although rarely mentioned in bibliometrics, multisets are actually often used. If one counts the number of citations received by a country, then a citing article may be counted several times as the set of all articles citing a given country is a multiset.

## 2.4 Definition of a small world

When authors write about the small-world property, it is not always clear which property they have in mind. Yet, most of the time they refer to the property we will refer to as SWA, meaning that the average distance between any two nodes is 'small'. This was done in the original article by Albert et al. (1999) and in many other ones such as (Girvan & Newman, 2002; Newman, 2003). But, for instance, Samoylenko et al. (2023) write that the small-world property means that the maximal distance between any two of a network's nodes scales logarithmically with its size, a property we will denote as SWD. Before going into details we point out that all the networks we will consider are assumed to be connected. They coincide with their so-called giant component. We assume given a sequence $(\Omega_N)_{N \in \mathbb{N}}$ of finite sets as introduced above, and a distance function d defined on $\mathbf{\Omega} = \bigcup_{N \in \mathbb{N}} \Omega_N$. We will use the term "small world" for a sequence $(\Omega_N)_{N \in \mathbb{N}}$ of finite sets satisfying the property SWD defined below.

2.4.1 Small worlds based on the diameter (SWD)

If $d_N, N \in \mathbb{N}$, is the diameter of $\Omega_N$, defined as

$$d_N = max\{d(A,B); \ A, B \ \epsilon \ \Omega_N \} \qquad (1)$$

then $(\mathbf{\Omega}, d)$ is a SWD if there exists a finite constant $C \geq 0$ such that

$$\lim_{N \to +\infty} \frac{d_N}{\ln(N)} = C \qquad (2)$$

Note that $d_N$ is short for diam($\Omega_N$). Stated otherwise: $(\mathbf{\Omega}, d)$ is a SWD if and only if the sequence $\left(\frac{d_N}{\ln(N)}\right)_{N \in \mathbb{N}}$ converges in $\mathbb{R}$.

2.4.2 Small worlds based on the average distance (SWA)

If $\mu_N, N \in \mathbb{N}$, denotes the average distance between two different elements in $\Omega_N$:

$$\mu_N = \frac{1}{N(N-1)} \sum_{\substack{A,B \ \in \ \Omega_N \\ A \neq B}} d(A,B) \qquad (3)$$

then $(\mathbf{\Omega}, d)$ is a SWA if there exists a finite number $C \geq 0$ such that

$$\lim_{N \to +\infty} \frac{\mu_N}{\ln(N)} = C \qquad (4)$$

2.4.3 Small worlds based on the median distance (SWMd)

If $Md_N, N \in \mathbb{N}$, denotes the median distance between two different elements in $\Omega_N$:

$$Md_N = median\{\{d(A,B); A, B \ \in \ \Omega_N, A \ \neq \ B \ \}\} \qquad (5)$$

then $(\mathbf{\Omega}, d)$ is a SWMd if there exists a finite number $C \geq 0$ such that

$$\lim_{N \to +\infty} \frac{Md_N}{\ln(N)} = C \qquad (6)$$

Note that for the SWMd case, we really need to use a multiset. For SWD and SWA, see definitions (1) and (3), it does not matter if we work within the context of a set, or a multiset.

It is obvious that if a sequence of finite sets is an SWD then it is also an SWA and an SWMd. If persons use the expression "six degrees of separation" they usually, but not always, refer to the SWA case. We also note that in calculating the average distance we did not include zero distances between a node and itself, but some authors, such as Newman (2003) do. For our purposes, this variant makes no difference.

2.5 An interpretation of the definitions of small worlds

Contrary to many practical investigations that focus on one specific case, we need to work with an infinite sequence of sets, as a small-world property is by definition a limiting property.

We concentrate on SWD, but the other types of small worlds can be interpreted similarly. There are two essentially different cases, namely C > 0 and C=0. If C > 0 then $d_N$ grows like *ln(N)*, which is the classical interpretation of a small world. If C=0 then we say that we have an ultra-small world (denoted as USWD, USWA, USWMd). This happens, e.g., if $d_N$ grows like *ln(ln(N))* or when there is an absolute bound on the sequence $(d_N)_{N \in \mathbb{N}}$ .

If all sets $\Omega_N$ are the same, and hence equal to **$\Omega$** , then automatically C = 0, and we have an ultra-small world.

## 2.6. Examples

2.6.1. The simplest case of a small world is the sequence of complete graphs. Here, each diameter is one, as is the average distance between any two nodes and the median distance. Hence, in each case C=0, and we have a USWD.

2.6.2 Another simple case of a USWD, and hence of an SWD, is the sequence of stars with one center and N-1 other nodes linked to the center at a distance of 1. The diameter is always equal to 2 and the average distance between two nodes is $\frac{2(N-1)}{N}$ (see appendix). The median distance is 1 for N=1,2,3, it is 1.5 for N= 4, and 2 if N > 4. In the three cases, the limiting value C is zero.

2.6.3. A chain of length N, with a distance of one between two consecutive nodes has a diameter of N-1. Then $\lim_{N \to +\infty} \frac{N-1}{\ln(N)} = +\infty$ and hence, this sequence is not an SWD. The average distance between two nodes is (N+1)/3. It can be shown that the median for a chain of length N (large N) is approximately $N\left(1 - \frac{\sqrt{2}}{2}\right) < \frac{N+1}{3}$ (Egghe, 2024). Limits for N → +∞ are in both cases equal to +∞, so we do not have a small world.

2.6.4. The well-known Poisson random graph of Erdös-Rényi (1959) (using the terminology of Newman (2003)) is a SWA with $\mu_N = \frac{\ln(N)}{\ln(z)}$ where z denotes the fixed mean degree of a node (taken from (Newman, 2003) who refers to (Bollobás, 2001)). Hence $\lim_{N \to +\infty} \frac{\mu_N}{\ln(N)} = \lim_{N \to +\infty} \frac{1}{\ln(z)}$ which is a constant.

## 2.6.5. Scale-free networks

Consider the following form of preferential attachment, also known as success-breeds-success, as introduced by Barabási and Albert (1999), leading to a scale-free network. Vertices are added to an existing network (starting with a network of v vertices) and joined to a fixed number m (2 ≤ m ≤ v) of earlier vertices, where each earlier vertex is chosen with probability proportional to its degree. Then Bollobás and Riordan (2004) showed that its diameter grows as $\frac{\ln(n)}{\ln(\ln(n))}$ where n is the number of steps. Consequently, such a network is a USWD. If m =1 then the diameter grows as ln(n), and thus the network is an SWD but not a USWD (Pittel, 1994; Bollobás & Riordan, 2004). Other constructions of scale-free networks may or may not lead to a small world, as shown e.g. in (Rozenfeld & ben-Avraham, 2007), who even obtain fractal scale-free networks.

3. A short review of studies of the small-world phenomenon

The first person to actually use the expression "six degrees of separation" is the American playwright John Guare. Indeed, neither Karinthy nor Milgram used the exact phrase "six degrees of separation." It was John Guare (1991) who originated the use of the term in his play, titled "Six Degrees of Separation".

One may say that the study of small worlds started with Erdös and Rényi (1959) and their random graph model. Yet, these models do not show any clustering as do most real-world networks. The situation changed considerably when Watts and Strogatz (1998) introduced their rewiring model, followed somewhat later by the preferential attachment growth model by Albert and Barabási (2002).

In 1999, Albert, Jeong, and Barabási (1999) found that in an N-node part of the Web, the average distance between two nodes was 0.35 + 2log$_{10}$(N). Given that they estimated the Web at that time to contain somewhat more than 800 million nodes, that yielded 19 degrees of separation (rounded).

As described by Barabási in his book *Linked* (2002), they, actually Jeong, started from a map of the nd.edu domain (where nd stands for the University of Notre Dame) domain by mapping about 300,000 documents within the University of Notre Dame, leading to, on average, 11 steps of separation. Then in a next series of steps they systematically started from a small portion of the Web, first with only 1,000 nodes, and calculated the separation between any two nodes on this sample. Next, they took a slightly larger piece, with 10,000 nodes, and determined the separation again. Repeating this for the largest systems their computer allowed they tried to find a trend in the obtained node-to-node distances, leading to the formula mentioned above. This method is called finite-size scaling. In this way, they could predict the separation on the whole Web. At that time, namely 1998, the size of the publicly indexable Web was estimated to be around 800 million nodes and hence they predicted that the diameter of the Web was 18.59, close to 19. This proved that the WWW has a small-world property (Albert & Barabásì, 2002).

Subsequent measurements by Bröder et al. (2000) found that the average path length between nodes – if there existed one - in a 50-million node sample of the World Wide Web was 16, roughly in agreement with the finite size prediction for a sample of this size.

Since then, small-world properties have been observed in a wide range of networks, including investment banking (Baum, Shipilov, Rowley, 2003), strategic technology alliances (Verspagen & Duysters, 2004), the brain connectome (Bullmore & Bassett, 2011), airport networks (Guimerà & Amaral, 2004), connected proteins (Bork et al., 2004), keyword networks (Zhu et al., 2013) and patent citations (Hung & Wang, 2010).

We already noted that finding a path from one node in a complex network to another is often possible, but finding the shortest path is another matter. Indeed, in reality, one only has local information to decide on the next step to the target node. Kleinberg (2000) showed that under certain conditions efficient navigability is possible, but only some small world networks meet these conditions.

Generally speaking, later authors gave the most attention to scale-free networks, with a power law degree distribution. For these Bollobás et al., (2001) have shown that the average distance between nodes scales as $\frac{\ln(N)}{\ln(\ln(N))}$ showing that these networks are USWA.

An almost complete review of the first years of studies on complex networks, including random graphs and small worlds can be found in Newman (2003).

We end this short review by recalling that more recently also time-varying small-world networks have been studied e.g., in (Tang et al., 2010).

4. A construction of a network with the small world property

In this section, we perform a step-wise construction and show when the resulting sequence of sets leads to a small world. In this construction, we will require that at each step there are at least two elements in the disjoint union of the constructed sets, that have a distance equal to two times the number of performed steps.

Let $\omega_0 = \Omega_0 = \{b\}$, a set with one element, namely b. This is the base set. The number of elements in $\Omega_0$, denoted as $N_0$ is equal to 1. Next, we perform the first step. We take a set $\omega_1$ with $a_1 \geq 1$ elements (where we set $a_0 = 1$). The distance between each of these elements and *b* is set equal to one, while the distance between any two of the elements in $\omega_1$ is for the moment not defined. Note that we have at least two elements whose distance is equal to two (two times the number of steps performed). Now we denote $\omega_0 \cup \omega_1$ $as$ $\Omega_{N_1}$, with $N_1 = 1 + a_1$, i.e., the number of elements in $\omega_0 \cup \omega_1$ (technically speaking this is a disjoint union, but we just use the standard notation for a union).

We continue this construction: if we have already performed n-1 steps and hence have constructed $\Omega_{\sum_{i=0}^{n-1} a_i} = \Omega_{N_{n-1}} = \bigcup_{i=0}^{n-1} \omega_i$ , then we take a set $\omega_n$ with $a_n > 1$ elements. For each element in $\omega_n$ we set the distance to at least one element in $\omega_{n-1}$ equal to 1 (other distances are as yet undefined). Moreover, there must be two different elements in $\omega_n$ whose distance is defined with respect to two elements in $\omega_{n-1}$ who have a mutual distance of n-1. Recall that such two elements in $\omega_{n-1}$ exist by the inductive procedure. Then we set $\Omega_{\sum_{i=0}^{n} a_i} = \Omega_{N_n} = \bigcup_{i=0}^{n} \omega_i$. The index of the sets $\omega_i$ indicates the number of steps in this construction. Clearly $N_n = \sum_{i=0}^{n} a_i$ tends to infinity when n tends to infinity.

For A, B belonging to a general set $\Omega_N$ we define d(A,B) = d(B,A), equal to the shortest distance between A and B via the construction above. By construction, this distance function d satisfies the triangle inequality.

Notation. We denote the set of numbers $\{N_n\}$ obtained in this construction as **M**.

Lemma 1.

$$\forall n \in \mathbb{N}, \forall A, B \in \Omega_{N_n} : d(A,B) \leq 2n \tag{7}$$

$$\text{and } \exists A, B \in \omega_n \subset \Omega_{N_n} : d(A,B) = 2n \tag{8}$$

$$\text{hence } d_{N_n} = 2n \tag{9}$$

Proof. By induction.

For n=1 we have by the first step in the construction above that $d(A,B) \leq 2$ and if, $A, B \in \omega_1 : d(A,B) = 2$. Hence $d_{\Omega_1} = 2$.

Induction step. We assume that formulae (7) and (8) are shown for n-1 (n≥2) and show that formulae (7) and (8) hold for n.

Consider any $A, B \in \omega_n$, then we know that there exist points C and D in $\omega_{n-1}$ : $d(C,D) \leq 2(n-1)$, $d(A,C) \leq 1$ and $d(B,D) \leq 1$ . Hence, $d(A,B) \leq d(A,C) + d(C,D) + d(B,D) = 2n$. If $A, B \in \Omega_{N_n} \setminus \omega_n$ then we already know, by induction, that $d(A,B) \leq 2(n-1) < 2n$. This proves (7) for the case n. Moreover, we know that there exist two points $C_0, D_0 \in \omega_{n-1}$ and two points $A_0, B_0 \in \omega_n$ such that: $d(C_0, D_0) = 2(n-1)$, $d(A_0, C_0) = 1$ and $d(B_0, D_0) = 1$ . By construction $d(A_0, B_0) = 2n$ , which shows formulae (8) and (9) for the case n, i.e., $d_{N_n} = n$.

Theorem 1.

The sequence $\left(\Omega_{N_n}\right)_{n \in \mathbb{N}}$ as constructed above is an SWD

$$\Leftrightarrow$$

$$\exists\, a > 1 \text{ such that } \lim_{n \to \infty} \frac{n}{\log_a(\sum_{i=0}^{n} a_i)} \leq 1 \tag{10}$$

Formula (10) is, moreover, always satisfied if

$$\lim_{n\to\infty} \frac{a^n}{\sum_{i=0}^n a_i} \leq 1 \tag{11}$$

Proof. From Lemma 1 we know that

$$\sum_{i=0}^{\binom{d_{N_n}/2}} a_i = \sum_{i=0}^n a_i = N_n \tag{12}$$

Equation (12) is the most important relation concerning SWD. Indeed, defining

$$f: \mathbb{N} \to \boldsymbol{M}: m \to \sum_{i=0}^m a_i = N_m \tag{13}$$

and

$$g: \boldsymbol{M} \to \mathbb{N}: N_k \to \frac{d_{N_k}}{2} \tag{14}$$

We see that for all n, $(f{\circ}g)(N_n) = f\left(\frac{d_{N_n}}{2}\right) = \sum_{i=0}^{\frac{d_{N_n}}{2}} a_i = N_n$ and $(g{\circ}f)(n) = g(\sum_{i=0}^n a_i) = g(N_n) = \frac{d_{N_n}}{2} = n$. This shows that the functions f and g are each other's inverse.

Now we have by definition that the sequence $\left(\Omega_{N_n}\right)_{n\in\mathbb{N}}$ as constructed above, is an SWD if and only if there exists C ≥ 0 such that

$$\lim_{n\to+\infty} \frac{d_{N_n}}{\ln(N_n)} = C \ \ or \ \lim_{n\to+\infty} \frac{n}{\ln(N_n)} = \frac{C}{2} \ \ or \ \lim_{n\to+\infty} \frac{g(N_n)}{\ln(N_n)} = \frac{C}{2} \tag{15}$$

where we used that $N_n = \#(\Omega_{N_n})$. Now we set D = C/2. Then we have an SWD if and only if: there exists D ≥ 0 such that

$$\lim_{n\to+\infty} \frac{f^{-1}(N_n)}{\ln(N_n)} \leq D \Leftrightarrow \lim_{n\to+\infty} \frac{n}{\ln(N_n)} \leq D \tag{16}$$

We have to prove the existence of D > 0 (or D = 0) that meets the inequality above. If D > 0, this is equivalent to proving the existence of a > 1 with D = 1/ln(a) > 0. For the case D = 0, we take any a > 1.

If (11) holds then

$$\lim_{n\to\infty}(n - \log_a(\sum_{i=0}^n a_i)) \leq 0 \tag{17}$$

From (17) we derive (10) as follows:

$$\lim_{n\to\infty} \frac{n}{\log_a\left(\sum_{i=0}^n a_i\right)} = \lim_{n\to\infty} \frac{n - \log_a(\sum_{i=0}^n a_i) + \log_a(\sum_{i=0}^n a_i)}{\log_a\left(\sum_{i=0}^n a_i\right)} \leq 1$$

This proves (10). □

Corollary

The sequence $\left(\Omega_{N_n}\right)_{n\in\mathbb{N}}$ as constructed above is an SWD

$$\Leftrightarrow$$

$$\exists\, b > 1 \text{ such that } \lim_{n\to\infty} \frac{b^n}{\sum_{i=0}^n a_i} \leq 1 \qquad (18)$$

Proof. ($\Leftarrow$) This follows immediately from Theorem 1, with b = a

($\Rightarrow$) By (10) we know that for each b, $1 < b < a$ (recall that a > 1), $\log_b a > 1$ and hence

$$\lim_{n\to\infty} \frac{n}{\log_b(\sum_{i=0}^n a_i)} = \lim_{n\to\infty} \frac{n}{(\log_b(a))(\log_a(\sum_{i=0}^n a_i))} < 1 \qquad (19)$$

Hence, we see that there exists $n_0 \in \mathbb{N}$ such that $n > n_0$ implies that $\frac{n}{\log_b(\sum_{i=0}^n a_i)} < 1$, and hence $n < \log_b(\sum_{i=0}^n a_i)$. This shows that there exists b > 1 such that $\forall\, n \geq n_0 : b^n < \sum_{i=0}^n a_i$, leading to (18). □

## 5. Comments about Theorem 1 and its corollary.

In this section, we use the same notation as in section 4. Theorem 1 and its corollary show that, as b > 1, being SWD is equivalent to an exponential increase (or faster) of $\sum_{i=0}^n a_i$.

In the special case that $\lim_{n\to\infty} \frac{n}{\log_a(\sum_{i=0}^n a_i)} = 0$ then also $\lim_{n\to\infty} \frac{d_{N_n}/2}{\log_a(N_n)} = 0$ and hence also $\lim_{n\to\infty} \frac{d_{N_n}}{\ln(N_n)} = 0$. In this case, we have an ultrasmall world based on the diameter. This situation happens if $d_{N_n}$ grows as $\ln(\ln(N_n))$ or, in the stronger case that $d_{N_n}$ has a finite upper limit.

If, for an *a > 1* formula (10) leads to a value in ]0,1] then we find the 'normal' small world case. If, however, formula (10) leads for all *a* to a limit strictly larger than 1, then the network increase does not lead to a small world. So formulae (10) and (11) provide precisely when the numbers $(a_j)_j$ lead to a small world, of course, for our specific model network.

One may say that the occurrence of a small world of any kind is rather obvious. In the case of our model (10) and (11) are rather weak conditions. If we assume that there are links in the sets $\omega_n$ then it would become even 'easier' to have a small world.

## 6. Relations between SWD, SWA and SWMd

We already observed that SWD $\Rightarrow$ SWA and SWD $\Rightarrow$ SWMd hold trivially. Now we will provide examples that the converse relations do not hold.

The examples we will construct are "combinations" of growing stars and growing chains.

We consider a chain of p nodes, where the last node is the center of a star with q additional nodes. Hence we have a network of N = p+q nodes and p-1+q links. We recall that the distance between connected nodes is 1.

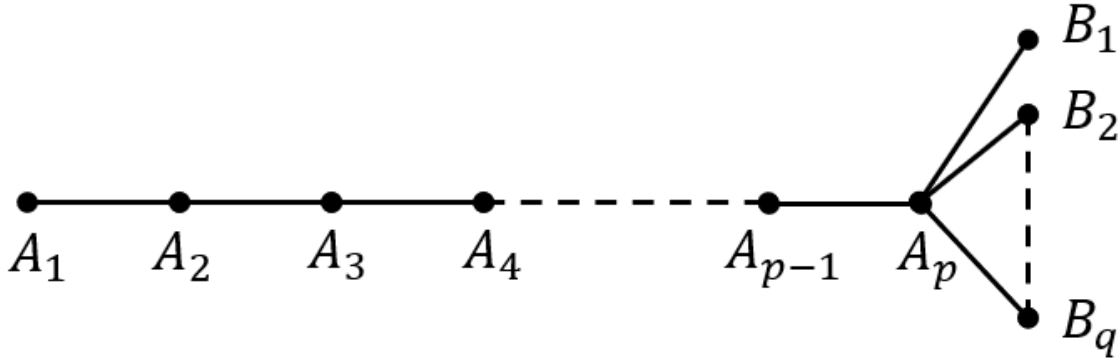

Fig.1. A network consisting of linear relations and a star

Let N = p+q be given as in Fig.1. The distance between every 2 B-points is 2, leading to a total of q(q-1). The total distance of the points in the A-chain is (p-1)p(p+1)/6. The total distance of one B-point to all A-points is p(p+1)/2. Hence the total distance of all B-points to all A-points is pq(p+1)/2. From this it follows that the total distance in the network is: q(q-1)+ (p-1)p(p+1)/6 + pq(p+1)/2. Hence the average distance, $\mu_N$, is:

$$(6q(q-1)+(p-1)p(p+1)+3pq(p+1))/(6(p+q)(p+q-1)/2) \qquad (20)$$

Now we take p = N' = $\lfloor \sqrt{N} \rfloor$, and q=N-N' . Then the average distance is

$\mu_N$ = (6(N-N')(N-N'-1)+(N'-1) N' (N'+1)+3N' (N-N' )(N'+1))/(3*N(N-1) )

which tends to 9/3=3 if N tends to infinity. Hence

$$\lim_{N \to +\infty} \frac{\mu_N}{\ln(N)} = 0 \qquad (21)$$

This proves that we even have an ultra-small world in distance. Now, clearly $d_N$ = p > $\sqrt{N} - 1$ and thus

$$\lim_{N \to \infty} \frac{d_N}{\ln(N)} \geq \lim_{N \to \infty} \frac{\sqrt{N} - 1}{\ln(N)} = \lim_{N \to \infty} \frac{\frac{1}{2\sqrt{N}}}{\frac{1}{N}} = \lim_{N \to \infty} \frac{N}{2\sqrt{N}} = +\infty$$

which shows that this network is not an SWD, and hence the property SWA does not imply SWD.

To show that SWMd does not imply SWD we adapt the parameters p and q from Fig.1. As a first step we take q such that

$$q(q-1) > (p+q)(p+q-1) - q(q-1) \qquad (22)$$

We require this inequality, because then the number of distances 2, which is q(q-1)+ q + (p-2) is larger than the total number of other distances, which is (p+q)(p+q-1) − q(q-1) − q − (p-2). Hence there are strictly more distances one or two than other distances, which shows that $Md_N \leq 2$. In that case, the network is an SWMd.

We see that condition (22) is equivalent with 2q(q-1) > N(N-1) which is, for instance, satisfied for q = 3N/4 and N > 4. In this case $d_N = p = N/4$, and because $\lim_{N \to \infty} \frac{d_N}{\ln(N)} = \lim_{N \to \infty} \frac{N/4}{\ln(n)} = +\infty$, this network is not an SWD.

Finally, we study the relation between SWA and SWMd.

Theorem 3

SWA ⇒ SWMd

Proof. This follows immediately from Markov's inequality which states that for a stochastic variable X with non-negative values and a > 0,

$$P(X \geq a) \leq \frac{E[X]}{a}$$

where P denotes a probability and E[X] is the average of X. Applying this inequality on an empirical finite sequence and taking the constant $a$ equal to the median value, yields $\frac{1}{2} \leq \frac{average}{median}$ , or the median of a finite sequence is smaller than or equal to twice the average (Chow & Teicher, 1978, p. 85). If now $\lim_{N \to +\infty} \frac{\mu_N}{\ln(N)} = C$, then $\lim_{N \to +\infty} \frac{Md_N}{\ln(N)} \leq \lim_{N \to +\infty} \frac{2\mu_N}{\ln(N)} \leq 2\,C$, which proves Theorem 3.

Finally, we show that the opposite implication does not hold.

Theorem 4

SWMd ⇏ SWA

Proof. We will use the same example as above with p = N/4. Then we know that SWMd holds and that SWD does not hold. Next, we show that even SWA does not hold. We know that the average distance is

(6q(q-1)+(p-1)p(p+1)+3pq(p+1))/(6(p+q)(p+q-1)/2)

With p = N/4 and hence q = 3N/4, $\mu_N$ is:

$$\frac{6\frac{3N}{4}\left(\frac{3N}{4}-1\right)+\left(\frac{N}{4}-1\right)\frac{N}{4}\left(\frac{N}{4}+1\right)+3\frac{N}{4}\frac{3N}{4}\left(\frac{N}{4}+1\right)}{3N(N-1)}$$

Then $\lim_{N\to\infty}\frac{\mu_N}{\ln(N)}=\lim_{N\to\infty}\frac{10N^3/64}{\frac{3N^2}{\ln(N)}}=+\infty$, which shows that this network is not an SWA network.

## 7. The majorization property and small worlds

### 7.1 Definition. The majorization property (Hardy et al., 1934).

Let $X=(x_1,\ x_2,\dots,x_N)$ be an N-array with $x_j\in\mathbb{R}^+, j=1,\dots,N$. If X and X' are N-arrays, ranked in decreasing order, then X is majorized by X' (equivalently X'majorizes X), denoted as $X-<_L\ X'$, if

$$\sum_{j=1}^{k}x_j\ \leq\ \sum_{j=1}^{k}x_j'\ for\ k=1,\dots,N-1\ and\ \sum_{j=1}^{N}x_j=\sum_{j=1}^{N}x_j' \qquad (23)$$

The index L in $X-<_L\ X'$ refers to the fact that this order relation corresponds to the order relation between the corresponding Lorenz curves. We recall that the Lorenz curve of the array X is the curve in the plane obtained by the line segments connecting the origin (0,0) to the points $\left(\frac{k}{N},\frac{\sum_{j=1}^{k}x_j}{\sum_{j=1}^{k}x_j'}\right)$, k= 1,…,N. For k = N, the endpoint (1,1) is reached. One observes that X is majorized by X' $(X-<_L\ X')$ if and only if the Lorenz curve of X' is situated above the Lorenz curve of X. $\sum_{j=1}^{N-1}x_j=\ \sum_{j=1}^{N-1}x_j'$

### 7.2 Majorization applied to $\alpha-$arrays

We recall from (Egghe, 2024) that if $\alpha_j$,  j= 1,…, N-1, denotes the number of times distance j occurs in network G, then array $AF=(\alpha_1,\alpha_2,\dots,\alpha_{N-1})$ is called the $\alpha-$array of the network G.

We will now apply majorization to $\alpha$-arrays AF and AF'. For an N-node network, the number N in the definition of majorization above becomes N-1 and all $\alpha$-values are natural numbers. Moreover, $\sum_{j=1}^{N-1}\alpha_j=\ \sum_{j=1}^{N-1}\alpha_j'=\frac{N(N-1)}{2}$. Furthermore, we will simply write -< for -<$_L$.

Notations

Let AF be an α-array of an N-node network $\Omega_N$. Then the largest index i such that $\alpha_i\neq0$ is nothing but the diameter of this N-node network. The average distance between two nodes is $\frac{2}{N(N-1)}\sum_{i=1}^{N-1}i\alpha_i$ denoted as $\mu_N$. The multiset of distances in this network contains $\frac{(N-1)N}{2}$ numbers and its median, $Md_N$ is either a natural number m or m-0.5.

Theorem 5

Let AF and AF' be $\alpha$-arrays of two networks $\Omega_N$ and $\Omega_N'$ both with N nodes. If AF -< AF' then

a) $d_N \geq d_N'$, where the accent refers to the network $\Omega_N'$ with AF' as α-array

b) $\mu_N \geq \mu_N'$,

c) $Md_N \geq (Md_N)'$

d) the opposite implications do not hold

Proof. a) Assume that $d_N < d_N'$. Then we know that there exists an index $i_0 \in \{2, \dots, N-1\}$ such that $\alpha_{i_0}' > 0$ and $\alpha_j = 0, \ j = i_0, \dots, N-1$. Then

$$\sum_{j=1}^{i_0-1} \alpha_j' < \sum_{j=1}^{i_0} \alpha_j' \leq \sum_{j=1}^{N-1} \alpha_j' = \frac{N(N-1)}{2} = \sum_{j=1}^{N-1} \alpha_j = \sum_{j=1}^{i_0-1} \alpha_j$$

which is in contradiction with the fact that AF -< AF'. Hence: $d_N \geq d_N'$.

b) The straightforward proof of this inequality has already been shown in (Egghe, 2024).

c) As AF -< AF' we know that for every i=1, …, N-1: $\sum_{j=1}^{i} \alpha_j \leq \sum_{j=1}^{i} \alpha_j'$. It follows then immediately that $Md_N \geq (Md_N)'$.

d) We provide two networks for which, $d_N > d_N'$, $\mu_N > \mu_N'$, and $Md_N = (Md_N)'$, but which are not comparable in the majorization order.

Consider Fig. 2

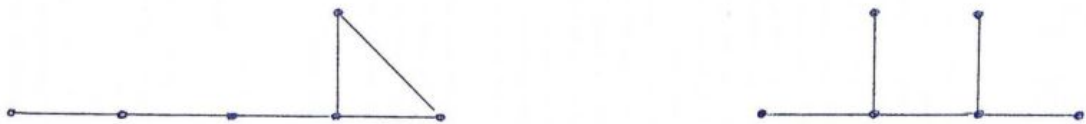

Fig.2. Two networks used in the proof of Theorem 5, part d)

For the network on the left-hand side in Fig.3 we have N = 6, AF = (6,4,3,2,0), $d_6 = 4$; $\mu_6 = 31/15$ and $Md_6 = 2$.

For the network on the right-hand side, we have N = 6, AF'= (5,6,4,0,0), $d_6' = 3$, $\mu_6' = 29/15$ and $Md_6' = 2$. This proves Theorem 5.

Corollaries

Consider the sequences $(\Omega_n)_{n \in \mathbb{N}}$ and $(\Omega'_n)_{n \in \mathbb{N}}$ of networks, such that for each $n \in \mathbb{N}$, $\#\Omega_n = \#\Omega'_n$. If now, there exists $n_0 \in \mathbb{N}$, such that for each $n \geq n_0$ : AF$_n$ -< AF'$_n$, where AF$_n$, resp. AF'$_n$ denote the $\alpha$-arrays of $\Omega_n$, resp. $\Omega'_n$ then

a) $(\Omega_n)_{n \in \mathbb{N}}$ is SWD implies that $(\Omega'_n)_{n \in \mathbb{N}}$ is SWD

b) $(\Omega_n)_{n \in \mathbb{N}}$ is SWA implies that $(\Omega'_n)_{n \in \mathbb{N}}$ is SWA

c) $(\Omega_n)_{n \in \mathbb{N}}$ is SWMd implies that $(\Omega'_n)_{n \in \mathbb{N}}$ is SWMd

d) the opposite relations do not hold.

These results follow immediately from Theorem 5.

Remark

Based on theorem 5 and its corollaries the following definition is acceptable and explained: For two networks X and X' with alpha sequences AF(X) and AF(X') we say that X' is a smaller world than X if AF(X) -< AF(X').

# 8. Conclusion

We recall that when studying small worlds it is essential to consider an infinite number of sets, as the notion of a small world implies a limiting property.

We introduced three types of small worlds: small worlds based on the diameter of the network (SWD), those based on the average geodesic distance between nodes (SWA), and those based on the median geodesic distance (SWMd). The exact hierarchical relation between these three types has been shown and is illustrated in Fig. 3. The well-known cases of a random network in the sense of Erdös-Rényi and the scale-free network as constructed by Barabási and Albert, but not necessarily scale-free networks constructed in another way, are both SWAs and SWMds.

Introducing the alpha-array of a network led us to a relationship between the majorization property between alpha-arrays and small-world properties.

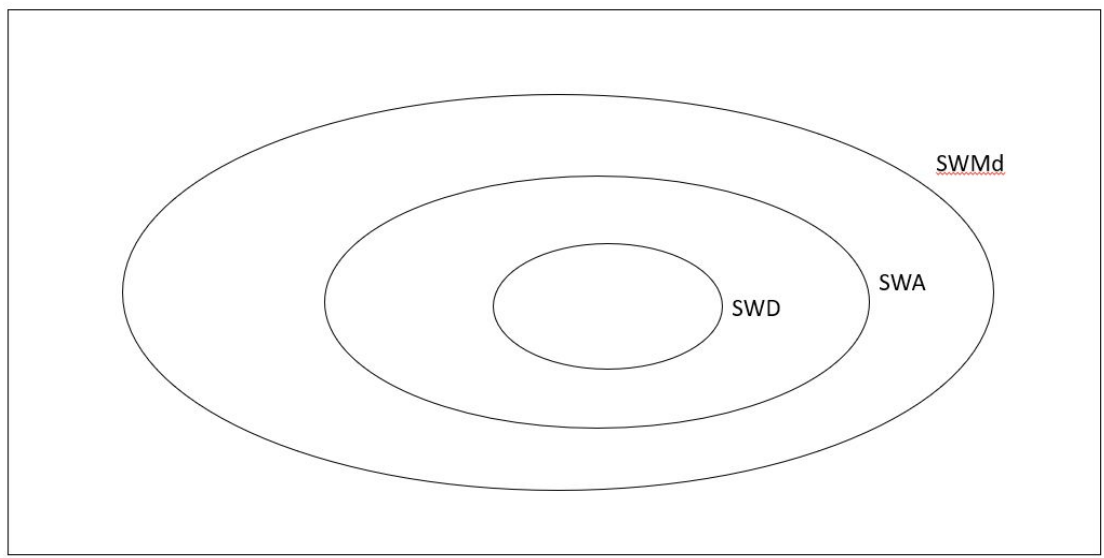

Figure 3. The relation between three types of small-worlds

Appendix. Average distance between two nodes in an N-node star

The distance from the center to any of the terminal nodes is 1, leading to a total distance of N-1. The distance of a terminal node to the center is 1, while the distance to any other node is 2, leading to a total of 1 + 2(N-2). As there are N-1 terminal nodes we obtain a total sum of distances equal to (N-1) + (N-1)(1 + 2(N-2)) = 2(N-1)$^2$.

Hence the average distance between any two nodes in an N-node star is $\frac{2(N-1)^2}{N(N-1)} = \frac{2(N-1)}{N}$.